\documentclass[prl,twocolumn,showpacs,superscriptaddress]{revtex4}

\usepackage[dvips]{graphicx}
\usepackage{amsmath}

\newcommand{\NVO}{NaV$_2$O$_5$ }

\newcommand{\etal}{\textit{et al.~}}

\newcommand{\dH}{$\Delta H$ }

\begin{document}

\title{Unconventional anisotropic superexchange in $\alpha'$-\NVO}

\author{M.~V.~Eremin}
\affiliation{EP V, Center for Electronic Correlations and Magnetism,
University of Augsburg, 86135 Augsburg, Germany}

\affiliation{Kazan State University, 420008 Kazan, Russia}

\author{D.~V.~Zakharov\footnote{Corresponding author:
Dmitri.Zakharov@physik.uni-augsburg.de}}

\affiliation{EP V, Center for Electronic Correlations and Magnetism,
University of Augsburg, 86135 Augsburg, Germany}

\affiliation{Kazan State University, 420008 Kazan, Russia}

\author{R.~M.~Eremina}
\affiliation{EP V, Center for Electronic Correlations and Magnetism,
University of Augsburg, 86135 Augsburg, Germany}

\affiliation{E. K. Zavoisky Physical Technical Institute, 420029
Kazan, Russia}

\author{J.~Deisenhofer}
\affiliation{EP V, Center for Electronic Correlations and Magnetism,
University of Augsburg, 86135 Augsburg, Germany}

\author{H.-A.~Krug~von~Nidda}

\affiliation{EP V, Center for Electronic Correlations and Magnetism,
University of Augsburg, 86135 Augsburg, Germany}

\author{G. Obermeier}
\affiliation{EP II, Institut f\"ur Physik, Universit\"{a}t Augsburg,
D-86135 Augsburg, Germany}

\author{S.~Horn}
\affiliation{EP II, Institut f\"ur Physik, Universit\"{a}t Augsburg,
D-86135 Augsburg, Germany}

\author{A.~Loidl}
\affiliation{EP V, Center for Electronic Correlations and Magnetism,
University of Augsburg, 86135 Augsburg, Germany}

\date{\today}

\begin{abstract}
The strong line broadening observed in electron spin resonance on
NaV$_2$O$_5$ is found to originate from an unusual type of the
symmetric anisotropic exchange interaction with
\textit{simultaneous} spin-orbit coupling on both sites. The
microscopically derived anisotropic exchange constant is almost two
orders of magnitude larger than the one obtained from conventional
estimations. Based on this result we systematically evaluate the
anisotropy of the ESR linewidth in terms of the symmetric
anisotropic exchange, only, and we find microscopic evidence for
precursor effects of the charge ordering already below 150~K.
\end{abstract}

\pacs{76.30.-v, 75.30.Et}

\maketitle



The isotropic exchange constants in one-dimensional antiferromagnets
are obtained by measurements of the magnetic susceptibility or by
inelastic neutron scattering. The anisotropic exchange
contributions, however, are only accessible by means of electron
spin resonance (ESR), because the spin-spin relaxation measured by
the ESR linewidth is driven primarily by the corresponding effective
local fields. Conventional theoretical estimations of anisotropic
exchange parameters yield values by far too small to describe the
experimental results. A prominent example for this problem is the
spin-ladder $\alpha'$-NaV$_2$O$_5$. This compound was initially
identified as a spin-Peierls system \cite{Isobe96} which triggered
enormous efforts to investigate the nature of this transition. It
was found that the system actually undergoes a charge-order (CO)
transition at $T_{\rm CO} \! \approx \! 34$\,K \cite{Ohama99} from a
uniform oxidation state of $V^{4.5+}$ ions at high temperature
\cite{Smolinski98} into a state with "zig-zag" type charge
distribution \cite{Seo96} accompanied by spin-gap formation.
Moreover, various experimental studies revealed an anomalous
behavior at about 200~K, far above $T_{\rm CO}$, which has been
attributed to the existence of charge fluctuations in the system
\cite{Damascelli00,Nishimoto98,Hemberger98}.

Previously, it was proposed that the spin relaxation in NaV$_2$O$_5$
is strongly affected by these charge fluctuations \cite{Lohmann00}.
ESR directly probes the spin of interest and, hence, is extremely
sensitive to such dynamic processes of the electronic structure. The
underlying mechanism of the spin relaxation, however, still remained
a matter of heavy debate \cite{Choukroun01,Oshikawa02}. In this
work, we identify the anisotropic exchange (AE) interaction as the
dominant source of line broadening and we calculate the AE
parameters on the basis of a microscopic charge-distribution
picture. With these parameters we are able to describe the angular
dependence of the ESR linewidth $\Delta H$ at temperatures above
$T_{\rm CO}$. The resulting temperature dependence of the exchange
parameters is a clear fingerprint of the increasing charge
fluctuations on approaching $T_{\rm CO}$.


All details concerning the preparation and characterization of the
crystals and the experimental ESR set-up have been published
previously \cite{Lohmann97}. The observed ESR signal in
$\alpha'$-NaV$_2$O$_5$ consists of a Lorentzian line with a
$g$-value $g \approx 2$, characteristic of a spin-only system with
quenched orbital moments \cite{Lohmann97}. The linewidth increases
monotonously from a value of 10~Oe at $T_{\rm CO}$ up to several
hundredth Oe above room temperature, with the linewidth $\Delta H_c$
for the magnetic field applied along the crystallographic $c$ axis
being about twice as large as $\Delta H_a$ and $\Delta H_b$ (along
the $a$ and $b$ axis).

\begin{figure}[tbp]
\centering
\includegraphics[width=85mm]{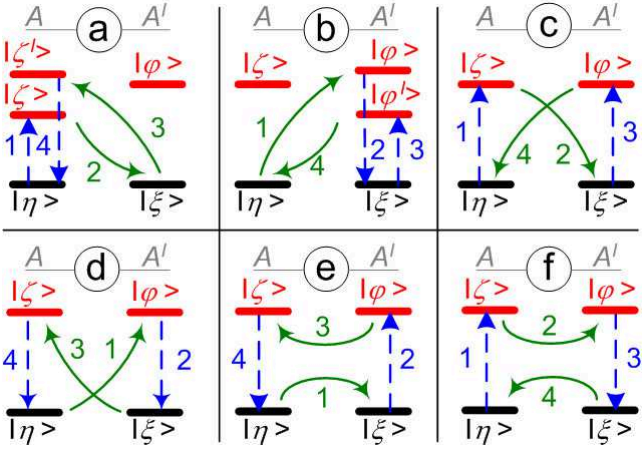}
\caption{Possible paths for AE between two sites $A$ (with ground
state $\vert \eta \rangle$, exited state $\vert \zeta \rangle$) and
$A'$ ($\vert \xi \rangle$ and $\vert \varphi \rangle$,
respectively). Solid arrows correspond to the effective hopping
integrals, dashed arrows indicate the matrix elements of the
spin-orbit coupling. Numeration corresponds to the sequence of the
matrix elements in forth order perturbation expansion (see
Eq.~\ref{Dab}).} \label{fig1}
\end{figure}

At first, let us consider possible origins for the line broadening
\dH in NaV$_2$O$_5$. Single-ion (\emph{S}=$1/2$), hyperfine and
spin-lattice relaxation were shown to be less important in
NaV$_2$O$_5$ (\cite{Yamada98,Zvyagin01,Hemberger98}). The
anisotropic Zeeman-effect is not relevant, because of nearly
equivalent $g$ tensors for all vanadium sites. Therefore, only three
sources remain to account for the broadening of the ESR spectra in
\NVO -- the dipole-dipole (DD), the symmetric anisotropic-exchange
(AE) and the antisymmetric Dzyaloshinsky-Moriya exchange (DM)
interactions. These contributions have been already estimated and
discussed in Ref.~\onlinecite{Yamada98}. The assumption that the DM
interaction is the main perturbation was based on the conventional
relation for the DM-vector $|d| \approx (\Delta g / g) |J|$ (where
$g$ and $\Delta g$ are the $g$ factor and its anisotropy,
respectively) \cite{Moriya60}. The ESR data could be modeled by this
approach, but it was necessary to assume the presence of strong
charge disproportions even at highest temperatures to allow for an
appropriate direction of the DM vector \cite{Lohmann00}. Later,
Choukroun \etal \cite{Choukroun01} questioned the dominance of the
DM interaction, showing that the contribution of the DM interaction
to the ESR linewidth in quantum-spin chains cannot be larger than
that of the AE. But the AE itself as taken from conventional
estimations is by far too small to account for the large linewidth
observed in NaV$_2$O$_5$. Therefore, such conventional estimations
have to be taken with care. A recent field-theoretical treatment of
quasi-one-dimensional $S=1/2$ antiferromagnetic chains came to
similar conclusions \cite{Oshikawa02}, because the DM interaction
was found to produce a divergence in the temperature dependence of
the linewidth $\Delta H_{\rm DM} \sim  T^{-2}$ for $T \ll J/k_{\rm
B}$. This is in contrast to the monotonic increase of $\Delta H$
with increasing temperature in NaV$_2$O$_5$ \cite{Lohmann00}. Such a
behavior, however, is in agreement with the theoretical expectation
for a dominant AE \cite{Oshikawa02}. Experimental investigations of
related compounds corroborate this expectation, too
\cite{KrugvNidda02,Eremina03,Demishev03,Zvyagin05}. In this respect
LiCuVO$_4$ received a key role \cite{KrugvNidda02}, because the DM
interaction can be completely ruled out by its crystal symmetry. The
linewidth is dominated by the AE, because the ring-exchange geometry
in the Cu-O$_2$ chains strongly enhances the AE as compared to the
conventional estimation $|J_{\rm AE}| \approx (\Delta g / g)^2 |J|$
\cite{Moriya60}. In the following we will provide detailed
microscopical estimations of this term in \NVO and show that the
angular and temperature dependencies of \dH can be completely
described in terms of this relaxation mechanism, only.

\begin{figure}[tbp]
\centering
\includegraphics[width=8cm]{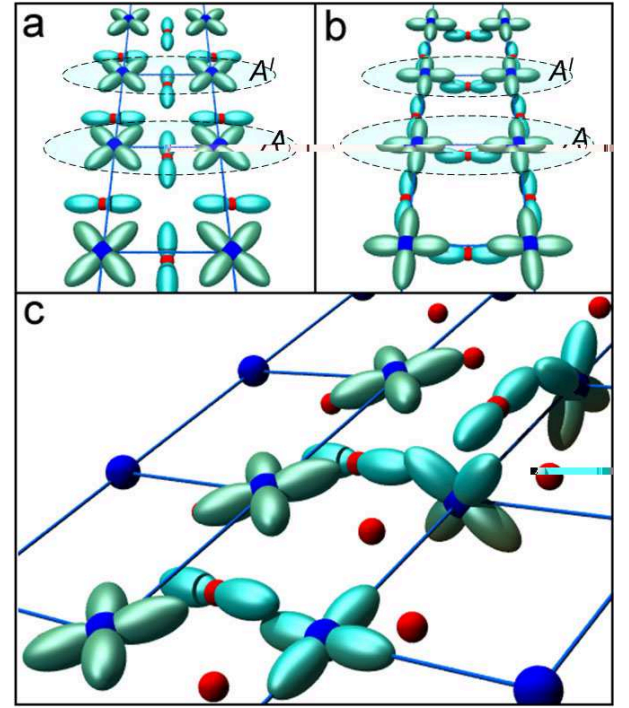}
\caption{Schematic pathway of AE between V ions in NaV$_2$O$_5$. Big
spheres denote V ions, small spheres O ions. Frame~(a): intra-ladder
exchange between the ground state $d_{xy}$ orbitals. Frame~(b):
intra-ladder exchange between the excited $d_{x^{2}-y^{2}}$ states.
Frame~(c): Possible inter-ladder exchange paths.} \label{fig2}
\end{figure}

Starting with the microscopic analysis of the AE paths, the
Hamiltonian for AE between two neighbouring sites $A$ and $A'$ can
be written as ${\cal H}_{AE} = S_A^\alpha \cdot D_{\alpha \beta
}^{AA'} \cdot S_{A'}^\beta$, where $\{\alpha, \beta\} = \{x,y,z\}$.
Taking into account all possible virtual processes (displayed
schematically in Fig.~\ref{fig1}) between site $A$ to site $A'$, we
derive the following expression for $D_{\alpha \beta }^{AA'}$ in
forth order of perturbation theory:
\begin{equation} \label{Dab}
\begin{array}{l}
D_{\alpha \beta }^{AA'}(\eta \xi ) = \frac{\lambda_{A}\lambda _{A'}}
{2 \Delta _{AA'}} \{ \frac{\langle\eta \vert l_{\alpha } \vert \zeta
\rangle} {\Delta_{\zeta \eta}} t_{\zeta \xi} t_{\xi
\zeta^\prime}\frac{\langle\zeta^\prime \vert l_{\beta} \vert \eta
\rangle} {\Delta_{\zeta^\prime \eta }}  + \\
+  t_{\eta \varphi } \frac{ \langle \varphi \vert l_{\alpha} \vert
\xi \rangle} {\Delta_{\varphi \xi}} \frac{ \langle \xi \vert
l_{\beta} \vert \varphi^\prime \rangle} {\Delta _{\varphi^\prime \xi
}} t_{\varphi^\prime \xi}+ \frac{\langle\eta \vert l_{\alpha }\vert
\zeta \rangle}{\Delta _{\zeta \eta }} t_{\zeta \xi }\frac{\langle\xi
\vert l_{\beta }\vert \varphi \rangle}{\Delta _{\varphi \xi
}}t_{\varphi \eta
}+ \\
+ t_{\eta \varphi }\frac{\langle\varphi \vert l_{\alpha }\vert \xi
\rangle}{\Delta _{\varphi \xi }}t_{\xi \zeta }\frac{\langle\zeta
\vert l_{\beta }\vert \eta \rangle}{\Delta _{\zeta \eta }} + t_{\eta
\xi } \frac{\langle\xi \vert l_{\alpha }\vert \varphi
\rangle}{\Delta _{\varphi \xi }}t_{\varphi \zeta }\frac{\langle\zeta
\vert l_{\beta }\vert
\eta \rangle}{ \Delta _{\zeta \eta }} + \\
+ \frac{\langle\eta \vert l_{\alpha }\vert \zeta \rangle}{\Delta
_{\zeta \eta }}t_{\zeta \varphi }\frac{\langle\varphi \vert l_{\beta
}\vert \xi \rangle}{\Delta _{\varphi \xi }} t_{\xi \eta }  \},
\end{array}
\end{equation} \\
where, e.g., $t_{\xi \zeta }$ is the effective hopping integral
between the states $\vert \xi \rangle$ and $\vert \zeta \rangle$ via
intermediate oxygens, and $\langle \xi \vert l_\alpha \vert \zeta
\rangle $ denotes the matrix element of the spin-orbit (SO) coupling
${\cal H}_{SO} = \lambda l_{\alpha} s_{\alpha} $. Here we assume
that the charge-transfer energy $\Delta _{AA'}$ from site $A$ to
site $A'$ is large compared to the crystal-field splittings $\Delta
_{cf} \equiv \Delta _{\zeta \eta }, \Delta _{\zeta ^{\prime }\eta
}$. The first two correspond to conventional AE processes
\cite{Bleaney52, Yosida96}, while the others (Fig.~\ref{fig1}(c-f))
and the general expression (\ref{Dab}) are presented to the best of
our knowledge for the first time. For example, in case (f) the
electron at site $A$ is excited via SO coupling from the ground
state $\vert \eta \rangle$ into the state $\vert \zeta \rangle$,
then it is transferred to the empty state $\vert \varphi \rangle$ at
site $A'$ and interacts via SO coupling with the electron in the
corresponding ground state $\vert \xi \rangle$. Finally, one of the
electrons hops from state $\vert \xi \rangle$ to the initial state
$\vert \eta \rangle$.

\begin{figure}[tbp]
\centering
\includegraphics[width=85mm]{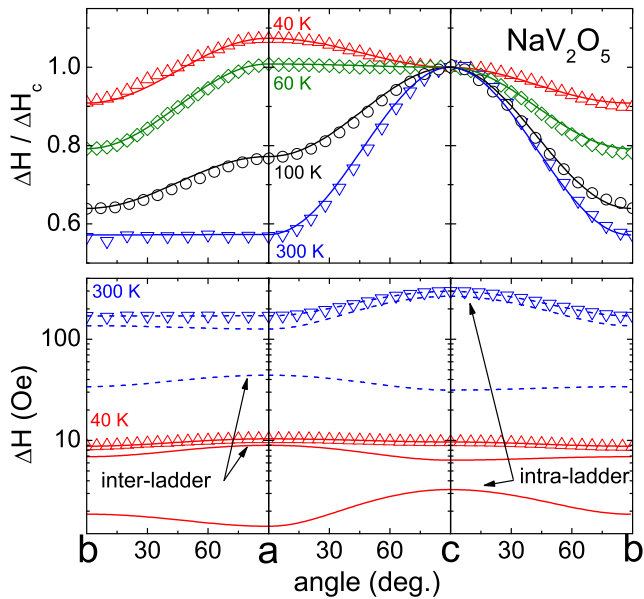}
\caption{Angular dependence of the ESR-linewidth at different
temperatures. Fit curves (lines) are described in the text. Upper
frame: normalized to the linewidth for the magnetic field applied
along the $c$ axis. Lower frame: illustration of the contributions
of intra- and inter-ladder AE to the linewidth far above (dashed
line) and near $T_{\rm CO}$ (solid line).} \label{fig3}
\end{figure}

Focusing now on NaV$_2$O$_5$, we recall that the electron is
distributed between two V-ions on the same rung. Correspondingly,
its ground state wave function $\vert \eta \rangle$ is given as a
superposition $c_{1} \, \vert d_{xy} \rangle - c_{2} \, \vert d_{xy}
\rangle$ of the two vanadium $d$-orbitals. Analogously, the ground
state of the electron on the adjacent rung is given by $\vert \xi
\rangle = c_{1}' \, \vert d_{xy} \rangle - c_{2}' \, \vert
d_{xy}\rangle$. We illustrated the corresponding $d$-orbitals
together with relevant bridging oxygen $p$-orbitals ($\pi$-bonding
with hopping integral $t_{\xi \eta }=t_\pi$) in Fig.~\ref{fig2}(a)
for the high-temperature limit, where all coefficients $c_1, c_2,
c_1', c_2'$ become equal to $1 / \sqrt{2}$. Note, that due to the
orthogonality of the wave functions processes (a)-(d)
(Fig.~\ref{fig1}) do not contribute to AE within one ladder in
NaV$_2$O$_5$. Therefore, we will now concentrate on processes (e)
and (f) and discuss the relevant excited states $\vert \zeta
\rangle$ and $\vert \varphi \rangle$ involved. Considering the
possible excitations of the electrons via SO coupling we find that
the largest contribution is obtained by the matrix element $\langle
d_{xy} \vert l_z \vert d_{x^{2}-y^{2}} \rangle = 2 i $. Hence, the
relevant excited states are the combinations $c_{1} \, \vert
d_{x^{2}-y^{2}} \rangle - c_{2} \, \vert d_{x^{2}-y^{2}}\rangle$ and
$c_{1}' \, \vert d_{x^{2}-y^{2}} \rangle - c_{2}' \, \vert
d_{x^{2}-y^{2}}\rangle$ for $A$ and $A'$ rungs, respectively. The
charge-distribution picture for the excited states ($\sigma$-bonding
via oxygen $p$-orbitals with hopping integral $t_{\zeta \varphi
}=t_{\sigma}'$) is shown in Fig.~\ref{fig2}(b).

Thus, using expression (\ref{Dab}) one can derive $D_{zz}$ as
\begin{equation}
D_{zz} = 8 \lambda^{2} \frac{t_{\pi} t_{\sigma}'}{\Delta_{cf}^{2}
\Delta_{AA'}}  [c_1^* c_1' + c_2^* c_2']^2~. \label{Dzz}
\end{equation}
To estimate $D_{zz}$ we use the free ion value $\lambda = 31$~meV
\cite{Abragam70}, the splitting between the $d_{xy}$ and the
$d_{x^{2}-y^{2}}$ states $\Delta _{cf} \simeq 0.36$~eV
\cite{Ohama97,Mazurenko02},  and the charge-transfer energy
$\Delta_{AA'} = 3$~eV \cite{Golubchiko97}. The hopping integral
$t_{\sigma}'$ cannot easily be calculated, however, one can assume
$t_{\sigma}' \approx t_{\pi} = 0.17$~eV \cite{Smolinski98} as a
lower bound for $t_{\sigma}'$, and we obtain $D_{zz} \approx
0.6$~meV in the high-temperature limit where the electron is equally
distributed on each rung. This yields a characteristic linewidth
$\Delta H \sim 300$~Oe in very good agreement with the experimental
linewidth. Note that our estimate is about two orders of magnitude
larger than previous results \cite{Yamada98}, corroborating the
importance of microscopic considerations for the estimation of AE
parameters.

Taking now into account possible inter-ladder exchange paths as
shown in Fig.~\ref{fig2}(c), we obtain contributions of comparable
strength for the inter-ladder AE. These paths involve a
$90^{\circ}$-exchange geometry, which has been discussed in detail
in Refs.~\onlinecite{Yushankhai99,KrugvNidda02,Tornow99}.
Considering all possible inter-ladder exchange paths in the
appropriate local coordinate systems \cite{Bencini90,Eremina03} we
find a maximal component of the effective AE tensor along the
crystallographic $a$ axis.

\begin{figure}[tbp]
\centering
\includegraphics[width=85mm]{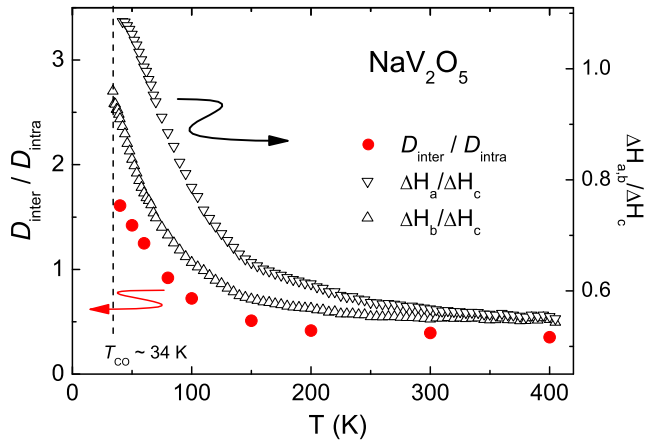}
\caption{Right ordinate: temperature dependence of the
linewidth-ratio for the magnetic field applied along $a$ or $b$ axis
normalized to $\Delta H_c$. Left ordinate: temperature dependence
ratio of the inter- to intra-ladder AE constants obtained from the
fitting of the angular dependencies of $\Delta H$.} \label{fig4}
\end{figure}

Having identified and estimated the source of the ESR line
broadening in NaV$_2$O$_5$, we will now apply this model to the
experimental data. The analysis of the angular dependencies in terms
of second moments has been discribed previously
\cite{KrugvNidda02,Eremina03}. The experimental angular dependencies
\cite{angular} of \dH are shown in Fig.~\ref{fig3} together with the
fit curves. As a result we derive the ratio of the two essential fit
parameters $D_{\rm intra}$ and $D_{\rm inter}$ for the AE parameters
within and between the ladders, respectively. Fig.~\ref{fig4} shows
the temperature dependence of the ratio $D_{\rm inter} / D_{\rm
intra}$ together with the linewidth ratios $\Delta H_{a}/\Delta
H_{c}$, $\Delta H_{b}/\Delta H_{c}$  (note that only the ratio of
the exchange parameters can be determined from the ESR linewidth at
temperatures $T < J/k_{\rm B}$ as discussed in Ref.
\onlinecite{Eremina03}). It can be clearly seen that at high
temperatures ($T > 150$~K) the dominant contribution to the line
broadening is given by the intra-ladder AE. On decreasing
temperature, below 150 K, the ratio strongly increases and the
inter-ladder contribution becomes dominant. This can be understood
taking into account the strong dependence of the AE parameters from
the coefficients $c_1, c_1', c_2, c_2'$ which describe the
electronic occupation on the vanadium sites. That means e.g. for
$D_{\rm intra} =D_{zz}$ (Eq.~\ref{Dzz}) the coefficient $[c_1^* c_1'
+ c_2^* c_2']$ is equal to 1 for the case $V^{4.5+} - V^{4.5+}$, and
vanishes for the "zig-zag" charge order ($V^{5+} - V^{4+}$) realized
below $T_{\rm CO}$ \cite{Seo96}. The observed increase of the
$D_{\rm inter}/D_{\rm intra}$ ratio already far above $T_{\rm CO}$
indicates that precursors of the developing CO set in at about
150~K, weakening considerably the intra-ladder AE.

Further evidence for the onset of charge disproportions above
$T_{\rm CO}$ has been reported by the strong frequency dependence of
the ESR linewidth between 34-100~K \cite{Nojiri2000}, and the
anomalous features observed by optical spectroscopy measurements
\cite{Nishimoto98, Smirnov1999, Damascelli00}. Note that
 the uniform susceptibility has significant deviations from the
Bonner-Fisher law already at $T < 200$~K, too \cite{Hemberger98}.
The coupling of these charge fluctuations to the lattice has been
revealed by a softening of the elastic constants below 100~K
detected by ultra-sound experiments \cite{Schwenk1999,Goto2003} and
by the shift of the phonon energy found in light-scattering
measurements around 80~K \cite{Fischer1999}. Moreover, we would like
to point out that similar observations in CuGeO$_3$ have been
explained in terms of lattice fluctuations existing already far
above the spin-Peierls transition \cite{Eremina03}. We believe that
the proposed spin-relaxation mechanism and the microscopic
estimations do not only apply for the case of NaV$_2$O$_5$, but may
allow to describe the spin dynamics in many transition-metal
compounds.

In summary, we have identified the symmetric anisotropic
super-exchange to be the source of the immense ESR line broadening
in NaV$_2$O$_5$. In this microscopic picture the dominant process
consists of the simultaneous virtual hopping of electrons between
the ground states and excited states of vanadium ions on neighboring
rungs of the ladder involving the spin-orbit coupling on both rungs.
This novel unconventional exchange process has not been considered
in the discussion of ESR line broadening before. The corresponding
AE parameter is found to be of the order of 1\% of the isotropic
exchange constant resulting in a high temperature limit of the ESR
linewidth of approximately $10^2$~Oe. On the basis of this
microscopic analysis we have shown that the ESR data can be entirely
described in terms of the symmetric anisotropic exchange only. The
temperature dependence of the linewidth and derived exchange
parameters evidences the presence of charge fluctuations in \NVO up
to 150~K on a microscopic level.

This work was supported by the German BMBF under Contract No.
VDI/EKM 13N6917, by the DFG within SFB 484 (Augsburg), by the RFBR
(Grant No. 03-02-17430) and RBHE (REC-007). One of us (D.~V.~Z.) was
supported by DAAD.



\begin{thebibliography}{99}
\bibitem{angular} The detailed reinvestigation of a series of samples has shown
that in best crystals (which have the smallest linewidth) the
additional modulation of the angular dependence of \dH within the
$(bc)$-plane reported in Ref.~\onlinecite{Lohmann00} is absent.
\end{thebibliography}
\end{document}